\documentclass[conference]{IEEEtran}
\IEEEoverridecommandlockouts
\usepackage{cite}
\usepackage{amsmath,amssymb,amsfonts}
\usepackage{algorithmic}
\usepackage{graphicx}
\usepackage{textcomp}
\usepackage{xcolor}
\def\BibTeX{{\rm B\kern-.05em{\sc i\kern-.025em b}\kern-.08em
    T\kern-.1667em\lower.7ex\hbox{E}\kern-.125emX}}

\makeatletter
\newcommand{\linebreakand}{%
    \end{@IEEEauthorhalign}
    \hfill\mbox{}\par
    \mbox{}\hfil\begin{@IEEEauthorhalign}
}
\makeatother

\begin{document}
	
\title{An Integrated Machine Learning and Deep Learning Framework for Credit Card Approval Prediction
}

\author{\IEEEauthorblockN{Kejian Tong*\thanks{*Kejian Tong and Zonglin Han are co-first authors.}}
\IEEEauthorblockA{
\textit{Independent Researcher}\\
Mukilteo, USA \\
 tongcs2021@gmail.com}
 \and
 \IEEEauthorblockN{Zonglin Han}
\IEEEauthorblockA{\textit{University of California San Diego}\\
La Jolla, USA \\
zoh003@ucsd.edu}
\and
\IEEEauthorblockN{Yanxin Shen }
\IEEEauthorblockA{\textit{Independent Researcher}\\
Hangzhou, China \\
shenyxedu@gmail.com}
\linebreakand
\and
\IEEEauthorblockN{Yujian Long}
\IEEEauthorblockA{\textit{Independent Researcher}\\
Frisco, USA \\
longyujian@gmail.com}
\and
\IEEEauthorblockN{Yijing Wei}
\IEEEauthorblockA{\textit{Northwestern University}\\
Evanston, USA \\
 yijingwei2022@u.northwestern.edu}
}

\maketitle

\begin{abstract}
Credit scoring is vital in the financial industry, assessing the risk of lending to credit card applicants. Traditional credit scoring methods face challenges with large datasets and data imbalance between creditworthy and non-creditworthy applicants. This paper introduces an advanced machine learning and deep learning framework to improve the accuracy and reliability of credit card approval predictions. We utilized extensive datasets of user application records and credit history, implementing a comprehensive preprocessing strategy, feature engineering, and model integration. Our methodology combines neural networks with an ensemble of base models, including logistic regression, support vector machines, k-nearest neighbors, decision trees, random forests, and gradient boosting. The ensemble approach addresses data imbalance using Synthetic Minority Over-sampling Technique (SMOTE) and mitigates overfitting risks. Experimental results show that our integrated model surpasses traditional single-model approaches in precision, recall, F1-score, AUC, and Kappa, providing a robust and scalable solution for credit card approval predictions. This research underscores the potential of advanced machine learning techniques to transform credit risk assessment and financial decision-making.
        
\end{abstract}
	
	
\section{Introduction}
Credit scoring is a critical tool in the financial sector, allowing lenders to assess the risk associated with extending credit to applicants. Traditionally, methods such as Logistic Regression (LR) and Decision Trees (DT) have been employed for this purpose. However, these conventional approaches often struggle with large, high-dimensional datasets and data imbalance issues, leading to less accurate predictions.

Modern advancements in machine learning (ML) and deep learning (DL) offer new opportunities to enhance credit scoring systems. Techniques like Support Vector Machines (SVM), k-Nearest Neighbors (KNN), and ensemble methods such as Random Forests (RF) and Gradient Boosting (GB) provide better handling of complex data and can offer significant improvements over traditional models. Deep learning models, particularly neural networks (NN), further enhance predictive capabilities by capturing intricate patterns in the data.

A major challenge in credit scoring is the data imbalance between creditworthy and non-creditworthy applicants. This imbalance can result in biased models that do not accurately predict the minority class, leading to suboptimal decision-making. To address this, techniques such as the Synthetic Minority Over-sampling Technique (SMOTE) are used to balance the dataset, improving model accuracy and robustness.

This paper proposes an advanced credit scoring framework that combines ML and DL techniques to improve prediction accuracy and reliability. We utilize extensive datasets of user application records and credit history, applying comprehensive preprocessing, feature engineering, and an ensemble approach integrating multiple models, including a neural network.

The main contributions of this work include:

Comprehensive Data Preprocessing: We address missing values and data imbalance issues while performing robust feature engineering.
Advanced Model Integration: We integrate various ML models and a neural network to leverage their combined strengths.
Improved Predictive Performance: Our approach demonstrates superior results in key metrics such as precision, recall, and F1-score compared to traditional methods.
This research underscores the potential of combining ML and DL techniques to enhance credit risk assessment and improve financial decision-making processes.
        
\section{Related Work}

Credit scoring has been a focal point of financial risk management for decades, with various techniques and models developed to enhance the prediction accuracy of creditworthiness. Traditional approaches have laid the foundation, but the advent of machine learning (ML) and deep learning (DL) has revolutionized the field by introducing more sophisticated methods capable of handling large and complex datasets.

L. Thomas et al. \cite{thomas2017credit} provide a comprehensive review of traditional credit scoring techniques and their applications. Hand and Henley \cite{hand1997statistical} review statistical classification methods used in consumer credit scoring, highlighting their limitations. C. Cortes \cite{cortes1995support} introduces support vector machines and their application in binary classification tasks. He et al. \cite{he2023space} present a 3D tool for visualizing and managing space, time, and workforce in construction projects.

W.E. Henley and D.J. Hand \cite{henley1996ak} discuss the application of k-nearest neighbors in consumer credit risk assessment. L. Breiman \cite{breiman2001random} proposes the Random Forests algorithm as a robust ensemble method for classification tasks. A multi-task learning method for monitoring tool surface changes in ultrasonic metal welding. Journal of Manufacturing Systems \cite{chen2021multi}. I. Brown and C. Mues \cite{brown2012experimental} compare the performance of various classification algorithms on imbalanced credit scoring datasets.

I. Goodfellow et al. \cite{goodfellow2016deep} offer an extensive overview of deep learning techniques and their applications in various domains. Y. LeCun et al. \cite{lecun2015deep} review advancements in deep learning and their impact on various fields, including credit scoring. He et al. \cite{he2022exploit} present a tool for optimizing space–time–workforce tradeoffs in construction scheduling.

Y. Cao et al. \cite{cao2023financial} study demonstrates that the Baichuan2-7B model significantly outperforms traditional models in financial text sentiment analysis. S.J. Yen and Y.S. Lee \cite{yen2009cluster} explore under-sampling methods for handling imbalanced data in classification tasks.

H. He and E.A. Garcia \cite{he2009learning} provide an overview of techniques for learning from imbalanced datasets. Y. Sun et al. \cite{sun2007cost} discuss cost-sensitive boosting methods for improving classification of imbalanced data. B. Zhang et al. \cite{zhang2024review} review recent advancements in NLP for text sentiment analysis. C.F. Tsai and M.L. Chen \cite{tsai2010credit} explore the use of hybrid machine learning techniques for credit rating and risk assessment.

J. Schmidhuber \cite{schmidhuber2015deep} summarizes key concepts and developments in deep learning and their applications. Y. Bengio et al. \cite{bengio2013representation} review techniques for representation learning in neural networks and their implications for various tasks. He et al. \cite{he2024synthesizing} integrate ontology and graph neural networks to improve decision-making in US bridge preservation.

In summary, the literature demonstrates significant advancements in the application of ML and DL techniques for credit scoring, highlighting the importance of addressing data imbalance and feature complexity. This study aims to build on this body of work by proposing a robust and scalable model that integrates multiple ML and DL methods to enhance credit scoring accuracy.

\section{Methodology}

The methodology employed in this research involves a systematic approach to developing and integrating multiple machine learning (ML) and deep learning (DL) models for predicting credit card approval decisions\cite{sun2024rapid}. Our approach includes data preprocessing, feature engineering, model selection, and an advanced ensemble strategy with a focus on neural networks (NN) to achieve optimal predictive performance.

\subsection{Key Features Analysis}
\subsubsection{Customer Characteristics Analysis}
\begin{figure}[h]
    \centering
    \includegraphics[width=0.5\textwidth]{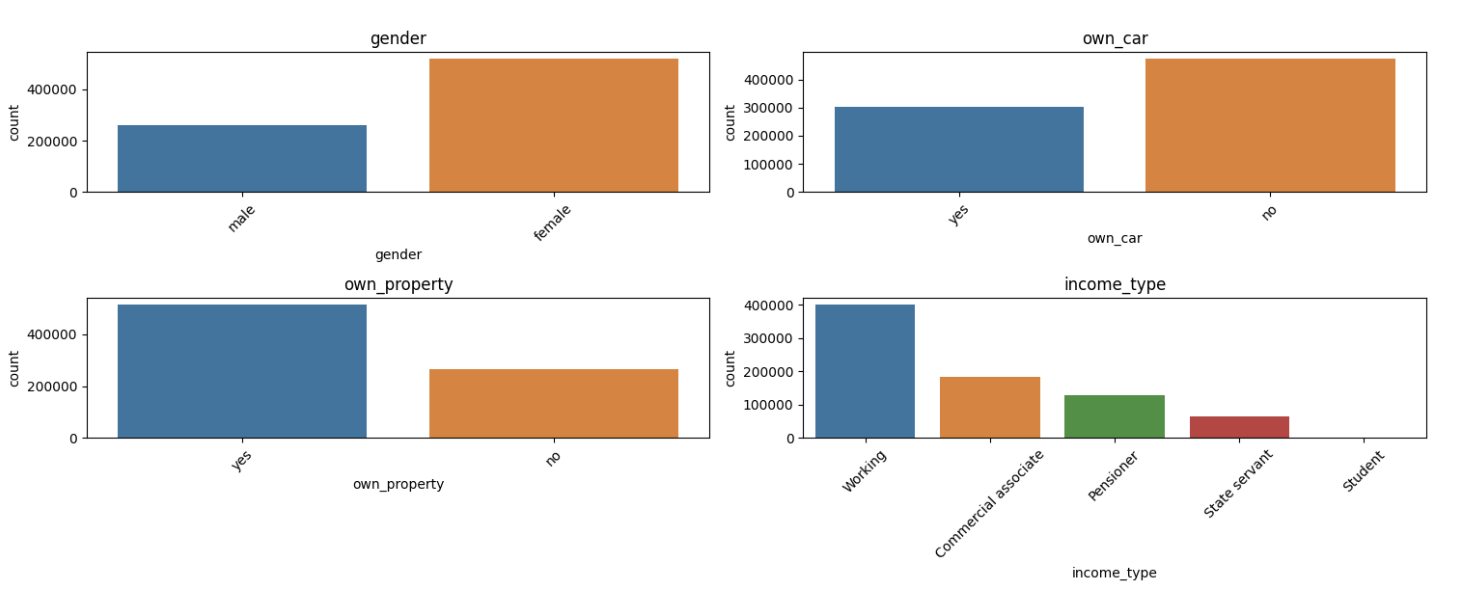}
    \caption{Comparison of customer characteristics by gender and asset ownership.}
    \label{fig:customer-characteristics}
\end{figure}

Our dataset shows a higher number of female customers compared to male customers, with more customers not owning a car and owning their homes. Females exhibit more overdue debts and bad loans, suggesting a potential need for gender-specific feature engineering, as illustrated in Figure~\ref{fig:customer-characteristics}.

\subsubsection{Loan Status Trends by Income Type and Gender}
\begin{figure}[h]
    \centering
    \includegraphics[width=0.4\textwidth]{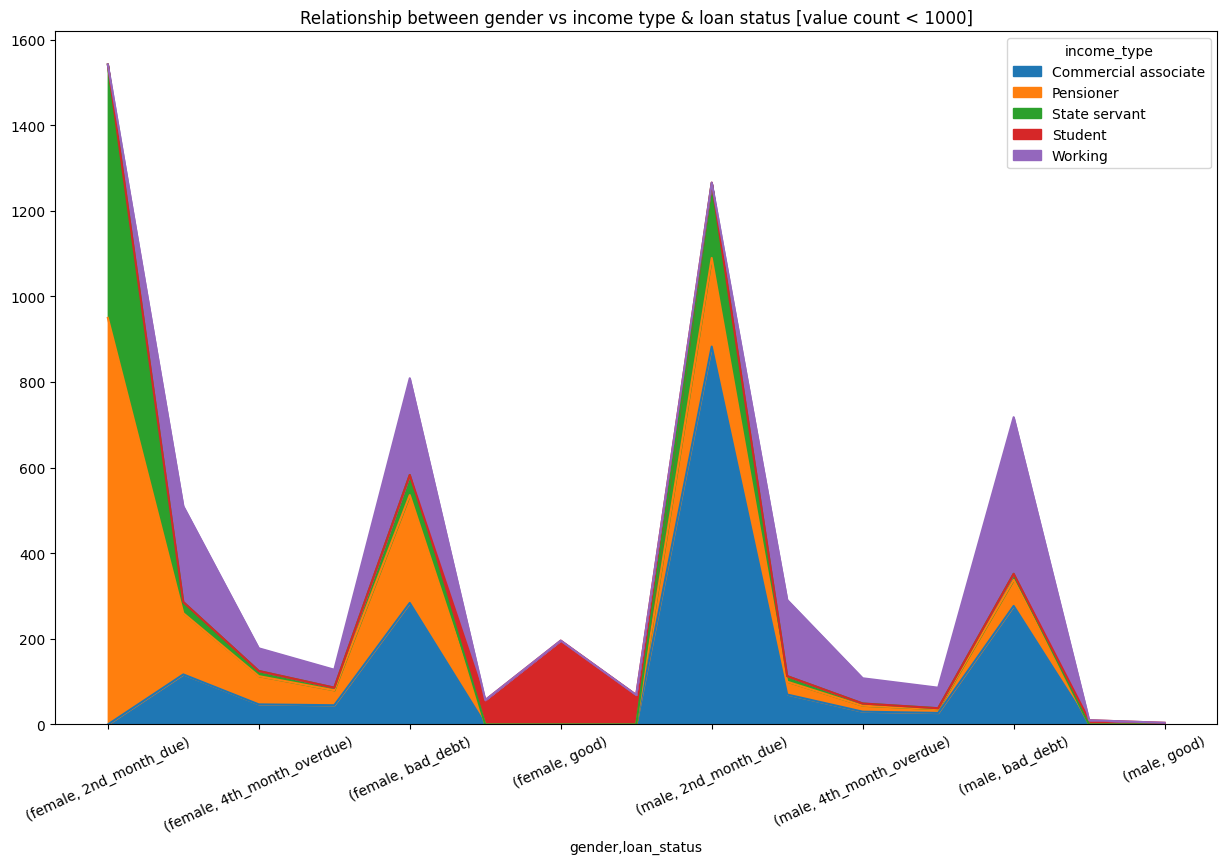}
    \caption{Loan status trends by income type and gender.}
    \label{fig:loan-status-trends}
\end{figure}

Male students generally do not have loans, and fewer male students have good loan statuses compared to females. Public servants, regardless of gender, have minimal overdue loans, indicating financial stability, as illustrated in Figure~\ref{fig:loan-status-trends}.

\subsubsection{Loan Status by Housing Type}
\begin{figure}[h]
    \centering
    \includegraphics[width=0.5\textwidth]{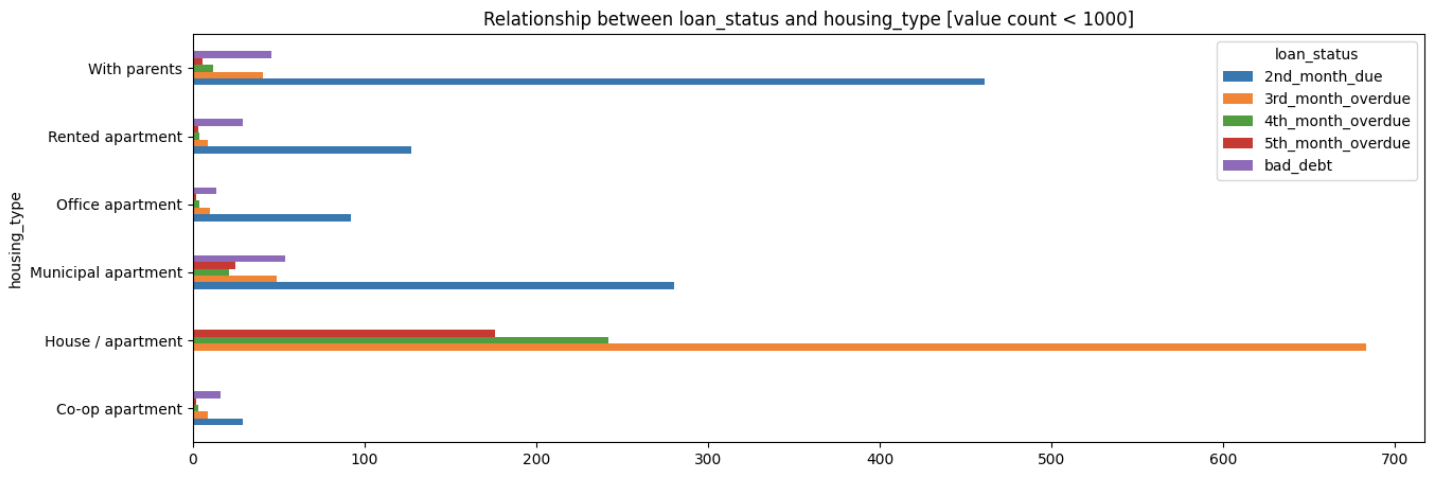}
    \caption{Loan status distribution by housing type.}
    \label{fig:loan-status-housing}
\end{figure}

Individuals owning their homes typically have good loan statuses or no loans. Those living with their parents have a higher proportion of loans maturing in the second month, impacting credit evaluation, as illustrated in Figure~\ref{fig:loan-status-housing}.

\subsection{Data Preprocessing}

Data preprocessing is a crucial step to ensure that the input data is clean, balanced, and suitable for model training. We utilized two primary datasets: user application records and credit history, which were merged to form a comprehensive dataset. The preprocessing steps include:

\subsubsection{Missing Value Handling}
\begin{figure}[h]
\centering
\includegraphics[width=0.35\textwidth]{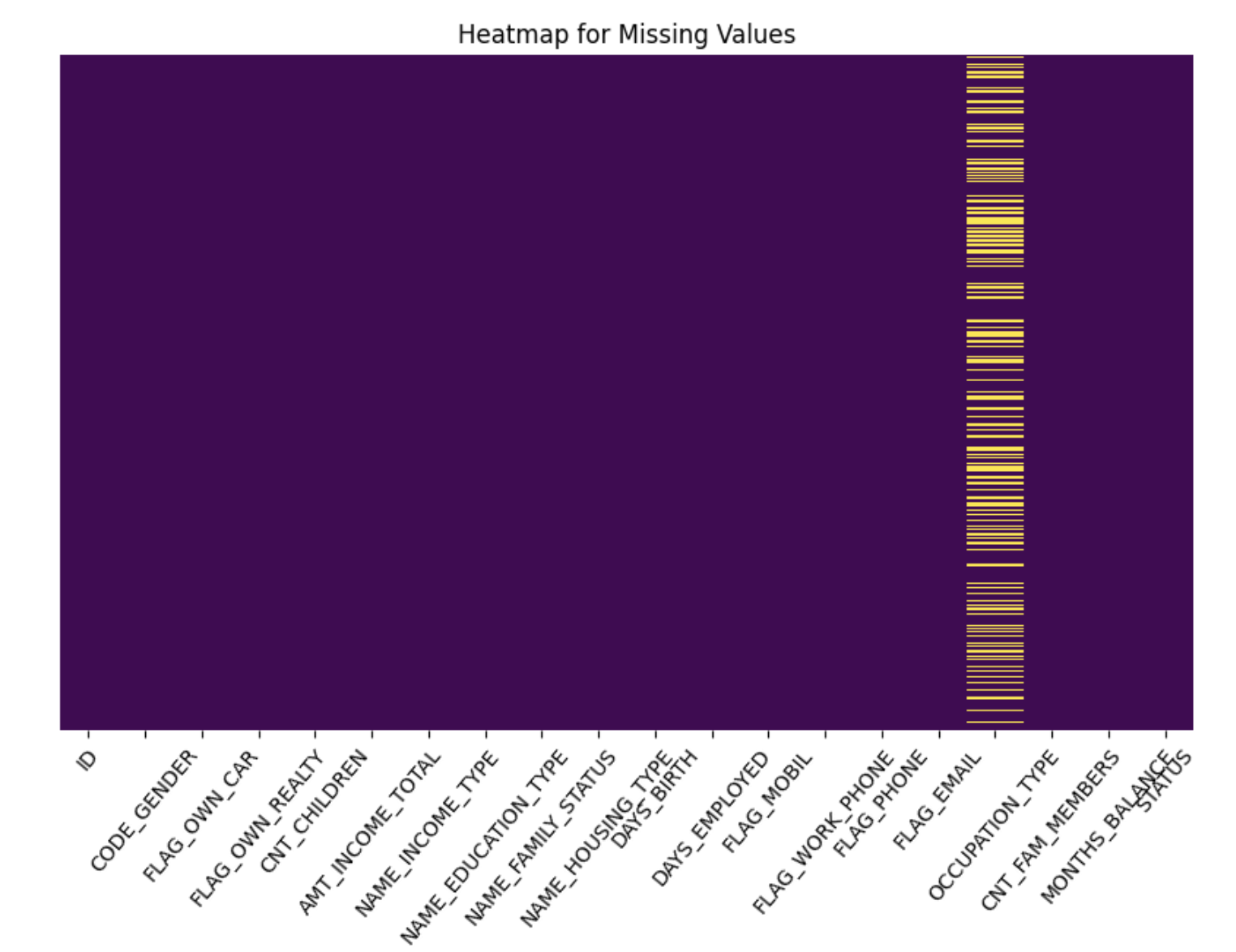}
\caption{Heatmap for Missing values}
\label{fig:Heatmap for Missing values}
\end{figure}

\begin{figure}[h]
\centering
\includegraphics[width=0.5\textwidth]{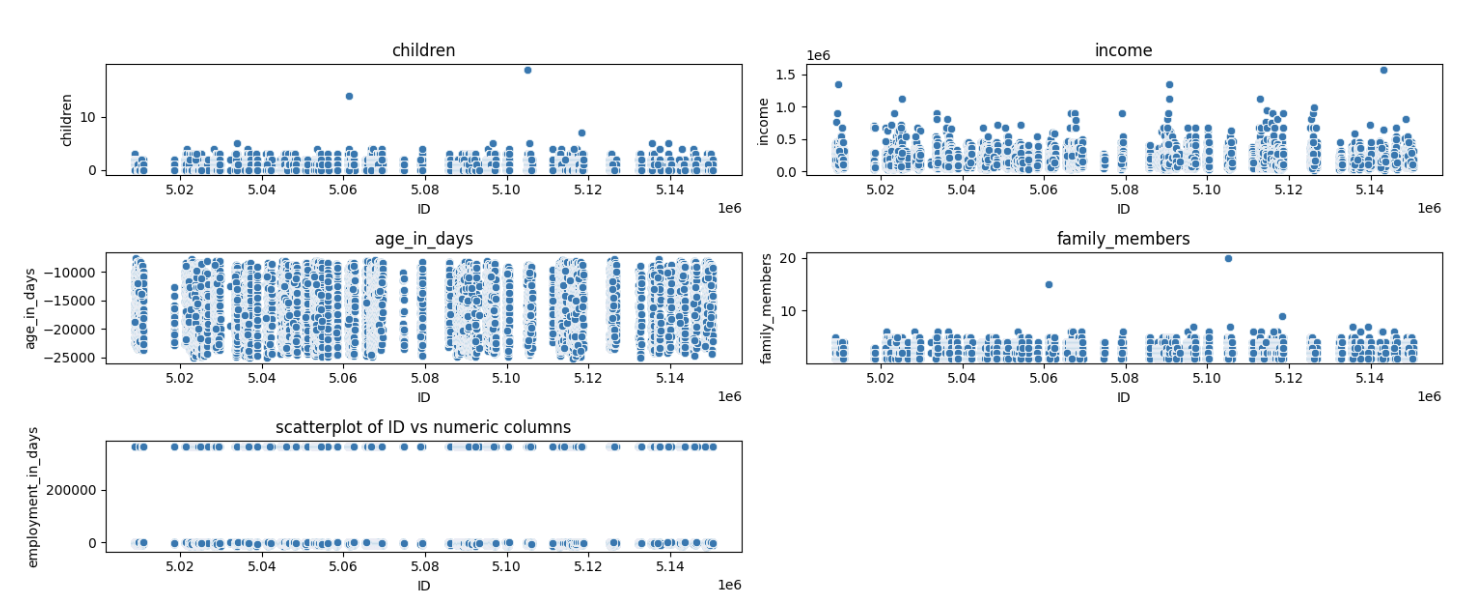}
\caption{Distribution of features}
\label{fig:Distribution of features}
\end{figure}

The dataset contained several missing values, notably in the OCCUPATION\_TYPE column, which had a missing rate of 30.86\%, as shown in Figure~\ref{fig:Heatmap for Missing values}. Given the significant amount of missing data, we decided to drop this column to maintain data integrity and avoid introducing biases in the model. For other columns with fewer missing values, imputation techniques were employed, using mean or median values for numerical features and the mode for categorical features,  as shown in Figure~\ref{fig:Distribution of features}. This approach helps to preserve the overall data distribution and minimize the potential for distortion.

\subsubsection{Data Merging}
The two datasets were merged using a common identifier (ID) to create a unified dataset with 777,715 samples. This merged dataset combined features from both user application records and credit history, providing a richer feature set for model training. The merging process involved careful alignment of records and handling any discrepancies to ensure data consistency.

\subsubsection{Feature Scaling}
Features were standardized to have zero mean and unit variance, which is essential for ensuring that the models, especially those relying on gradient-based optimization, perform optimally. The scaling was done using the formula:

\begin{equation}
x' = \frac{x - \mu}{\sigma}
\end{equation}

where \( x \) is the original feature value, \( \mu \) is the mean of the feature, and \( \sigma \) is the standard deviation. Standardization ensures that the different features contribute equally to the model's learning process, preventing dominance by any feature due to its scale.

\subsubsection{Encoding Categorical Variables}
Categorical variables were encoded using one-hot encoding, transforming each category into a binary vector. This approach ensures that categorical features are appropriately represented in the model without introducing ordinal relationships. One-hot encoding effectively transforms categorical data into a format suitable for input into ML algorithms, maintaining the categorical distinctions without implying any inherent order.

\subsection{Feature Engineering}

Feature engineering was conducted to enhance the dataset's predictive power by creating new features and transforming existing ones. This process involved the generation of new informative variables and the transformation of existing ones to better capture the underlying patterns in the data.

\subsubsection{Interaction Features}
We created interaction terms to capture the multiplicative effects between different features. For instance, the interaction between \texttt{Income} and \texttt{Credit History} was represented as:

\begin{equation}
\text{Interaction} = \text{Income} \times \text{Credit History}
\end{equation}

Interaction terms help in modeling the joint effect of multiple features, which can significantly impact the predictive power of the model by capturing complex relationships that are not evident from individual features alone.

\subsubsection{Polynomial Features}
Polynomial features up to the second degree were added to model non-linear relationships within the data. For example, for a feature \( x \), we included \( x^2 \). These polynomial transformations allow the model to capture non-linear interactions among features, enhancing its ability to fit complex data patterns.

\subsubsection{Temporal Features}
Temporal features such as Time Since Last Default were derived from the credit history data, providing a dynamic aspect to the creditworthiness prediction. These features capture the temporal dimensions of credit behavior, which are critical in understanding and predicting future credit risk.

\subsection{Model Development}

The model development phase involved selecting and training multiple ML models and a neural network, followed by integrating them into an ensemble framework. Each model was carefully selected and tuned to exploit its specific strengths and address its weaknesses, contributing to a robust ensemble.

\subsubsection{Base Models}
We employed several base models, each optimized for different aspects of the data:

\paragraph{Logistic Regression (LR)}
A fundamental model for binary classification, providing a probabilistic approach to prediction. The logistic regression model is defined as:

\begin{equation}
P(Y=1|X) = \frac{1}{1 + e^{-(\beta_0 + \beta_1 X_1 + \ldots + \beta_p X_p)}}
\end{equation}

where \( \beta_0 \) is the intercept, \( \beta_1 \ldots \beta_p \) are the coefficients for the predictors \( X_1 \ldots X_p \). Logistic regression is particularly valued for its interpretability and efficiency in handling large datasets.

\paragraph{Support Vector Machine (SVM)}
Effective for high-dimensional spaces, used with a linear kernel to separate the classes. The SVM optimization problem is formulated as:

\begin{equation}
\min_{w, b} \frac{1}{2} \|w\|^2 \quad \text{subject to} \quad y_i (w \cdot x_i + b) \geq 1, \forall i
\end{equation}

where \( w \) is the weight vector, \( b \) is the bias term, \( y_i \) is the class label, and \( x_i \) is the feature vector. SVMs are robust against overfitting, especially in high-dimensional feature spaces.

\paragraph{k-Nearest Neighbors (KNN)}
A non-parametric method that classifies based on the majority vote of the nearest neighbors. The prediction for a new data point \( x \) is given by:

\begin{equation}
\hat{f}(x) = \frac{1}{k} \sum_{i \in \text{kNN}(x)} y_i
\end{equation}

where \( k \) is the number of nearest neighbors, and \( y_i \) are the class labels of the neighbors. KNN is simple and effective for smaller datasets but computationally intensive for large datasets.

\paragraph{Decision Tree (DT)}
A tree-based model that splits the data into subsets based on feature values. The information gain for a feature \( X \) is calculated as:

\begin{equation}
H(X) = -\sum_{i=1}^n p(x_i) \log p(x_i)
\end{equation}

Decision trees are highly interpretable and capable of capturing non-linear relationships, but they are prone to overfitting.

\paragraph{Random Forest (RF)}
An ensemble of decision trees to improve robustness and accuracy. The prediction for an input \( x \) is the average of predictions from individual trees:

\begin{equation}
\hat{f}(x) = \frac{1}{B} \sum_{b=1}^B f_b(x)
\end{equation}

where \( B \) is the number of trees, and \( f_b(x) \) is the prediction from the \( b \)-th tree. Random forests reduce the variance and improve generalization by aggregating multiple decision trees.

\paragraph{Gradient Boosting (GB)}
A boosting method that builds an ensemble of weak learners to reduce bias and variance. The prediction function is updated iteratively:

\begin{equation}
F_m(x) = F_{m-1}(x) + \nu \sum_{i=1}^N \gamma_i I(x \in R_i)
\end{equation}

where \( F_m(x) \) is the updated prediction, \( \nu \) is the learning rate, \( \gamma_i \) are the weights, and \( I(x \in R_i) \) is the indicator function for the region \( R_i \). Gradient boosting focuses on correcting the errors of the previous models.

\paragraph{XGBoost}
An optimized version of gradient boosting with better performance and efficiency. The objective function for XGBoost is:

\begin{equation}
\min_{w} \sum_{i=1}^n \left[ L(y_i, \hat{y}_i) + \Omega(f_i) \right]
\end{equation}

where \( L \) is the loss function, \( \hat{y}_i \) is the predicted value, and \( \Omega \) is a regularization term. XGBoost incorporates advanced features like regularization, which helps prevent overfitting.

\subsubsection{Neural Network (NN)}
The NN model was designed with three hidden layers using ReLU activation functions and a sigmoid output layer for binary classification. The activation function for the hidden layers is:

\begin{equation}
\text{Activation: } f(x) = \max(0, x)
\end{equation}

and the output layer uses the sigmoid function for binary classification:

\begin{equation}
\text{Output: } P(Y=1|X) = \frac{1}{1 + e^{-z}}
\end{equation}

where \( z \) is the linear combination of the weights and inputs from the previous layer. Neural networks are powerful in capturing complex patterns in data due to their deep architecture.

\subsubsection{SMOTE for Data Imbalance}
To address data imbalance, SMOTE was applied to generate synthetic samples for the minority class, creating a balanced training dataset. The oversamping PR curve is shown in Figure~\ref{fig:PR curve}. The generation of new samples is defined as:

\begin{figure}[h]
\centering
\includegraphics[width=0.4\textwidth]{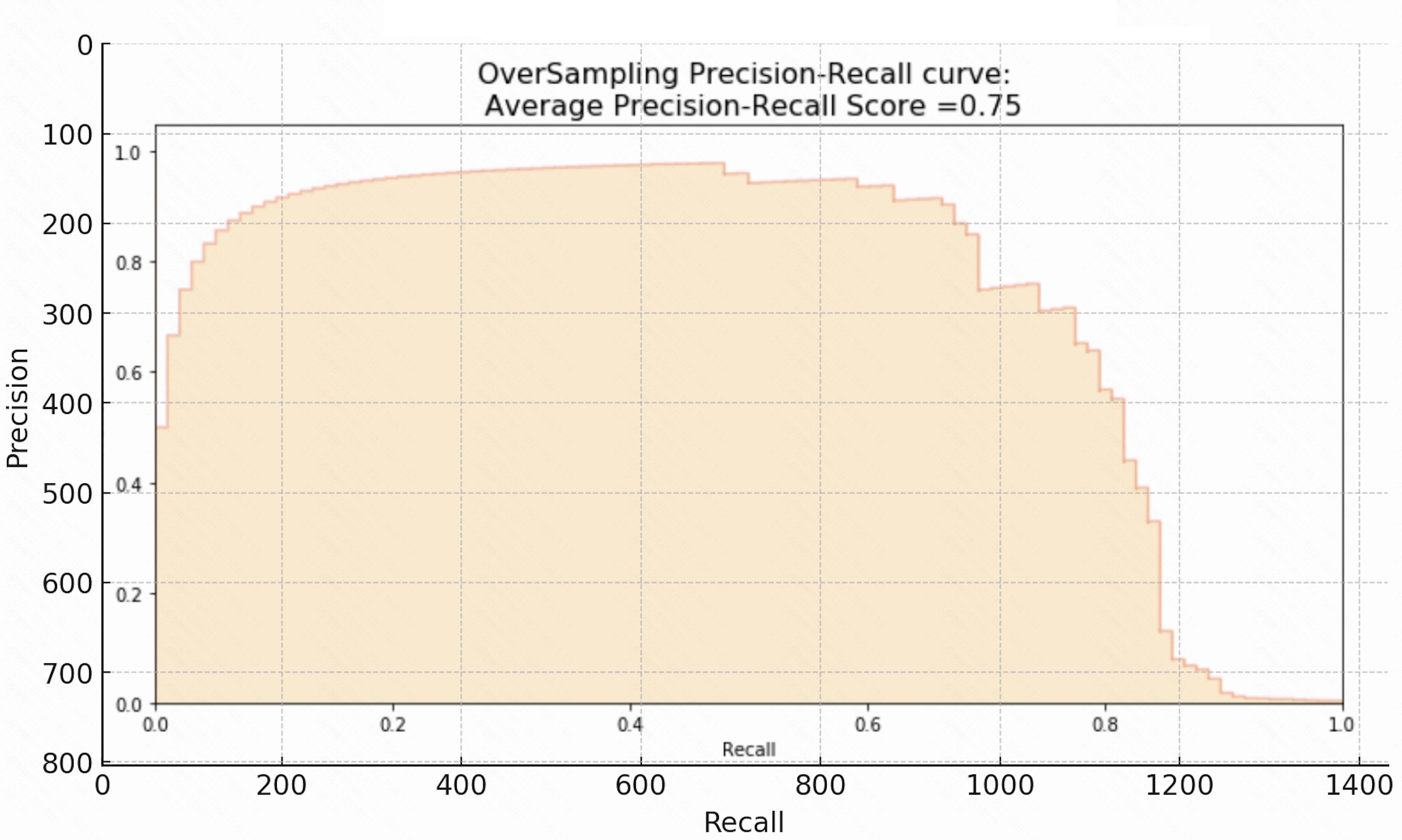}
\caption{Oversampling precision-recall curve}
\label{fig:PR curve}
\end{figure}

\begin{equation}
\text{Sample} = X_{minority} + \lambda \times (X_{neighbor} - X_{minority})
\end{equation}

where \( \lambda \) is a random number between 0 and 1, and \( X_{neighbor} \) is one of the nearest neighbors of \( X_{minority} \). SMOTE helps in reducing bias towards the majority class by providing a balanced dataset for training.

\subsection{Model Integration and Ensemble Strategy}

\begin{figure}[h]
\centering
\includegraphics[width=0.5\textwidth]{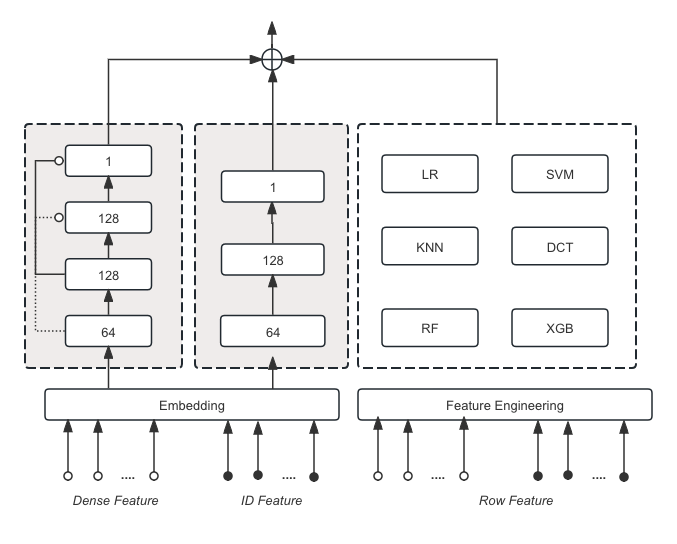}
\caption{Model architecture}
\label{fig:model}
\end{figure}

The final model combines predictions from multiple base models and the neural network using an ensemble strategy. This approach aims to leverage the strengths of individual models and reduce their weaknesses, resulting in a more robust and accurate overall prediction.  The overall architecture is shown in Figure~\ref{fig:model}.

\subsubsection{Stacking Ensemble}
In stacking, base models generate level-1 predictions, which are then used as inputs for a meta-learner model. The meta-learner, in this case, a neural network, combines these predictions to produce the final output. The stacked generalization is defined as:

\begin{equation}
\hat{Y} = g \left( h_1(X), h_2(X), \ldots, h_n(X) \right)
\end{equation}

where \( h_i(X) \) represents the predictions of the \( i \)-th base model, and \( g \) is the function learned by the meta-learner. Stacking leverages the diverse strengths of various models to enhance predictive performance.

\subsubsection{Neural Network Embedding Integration}
After the base models generate their predictions, these outputs, along with the input features, are processed through a neural network for embedding. The input features are divided into dense features and categorical ID features. The neural network embeddings are then integrated with the base models' outputs, enabling implicit learning from multiple dimensions of feature information. This combined approach facilitates comprehensive feature representation and enhances predictive accuracy.

\subsubsection{Model Training and Prediction}
Each base model was trained sequentially, and parameter optimization was performed using Optuna. For example, parameter tuning for XGBoost involved iterative searches to identify the optimal set of parameters, improving the model's predictive power. This systematic training and optimization ensure that each model contributes effectively to the final ensemble prediction.

\subsection{Model Evaluation}

The performance of the models was assessed using a variety of metrics to ensure a comprehensive evaluation. Precision, recall, and F1-score were used to evaluate the balance between positive predictions and their accuracy. Additionally, the Area Under the Curve (AUC) provided a measure of the models' ability to distinguish between classes, while Cohen’s Kappa offered insight into the agreement between predicted and actual classifications beyond chance.

\begin{equation}
\text{Precision} = \frac{TP}{TP + FP}
\end{equation}
where \(TP\) represents true positives and \(FP\) represents false positives.

\begin{equation}
\text{Recall} = \frac{TP}{TP + FN}
\end{equation}
where \(FN\) represents false negatives.

\begin{equation}
\text{F1-score} = \frac{2 \cdot \text{Precision} \cdot \text{Recall}}{\text{Precision} + \text{Recall}}
\end{equation}

\begin{equation}
\text{AUC} = \int_0^1 TPR(FPR) \, d(FPR)
\end{equation}
where \(TPR\) is the true positive rate and \(FPR\) is the false positive rate.

\begin{equation}
\text{Cohen's Kappa} = \frac{P_o - P_e}{1 - P_e}
\end{equation}
where \(P_o\) is the observed agreement and \(P_e\) is the expected agreement by chance.

\section{Experiments Results}

In this section, we present the experimental setup and results from evaluating various models, including the NN+Ensemble approach, for predicting credit card approval.

\subsection{Experimental Setup}

The dataset was divided into training (80\%) and testing (20\%) sets. We used 5-fold cross-validation on the training set to optimize hyperparameters and prevent overfitting. Model performance was evaluated using several key metrics: precision, recall, F1-score, AUC (Area Under the Curve), and Cohen’s Kappa.

\subsection{Model Performance}

Table~\ref{tab:base_models} summarizes the performance metrics for the individual base models and the NN+Ensemble approach. The NN+Ensemble model demonstrated superior performance across all evaluated metrics, indicating its effectiveness in capturing complex relationships and addressing data imbalance.

\begin{table}[htbp]
\centering
\caption{Performance metrics for various models.}
\label{tab:base_models}
\begin{tabular}{|c|c|c|c|c|c|}
\hline
\textbf{Model} & \textbf{Precision} & \textbf{Recall} & \textbf{F1-Score} & \textbf{AUC} & \textbf{Kappa} \\
\hline
LR & 0.61 & 0.61 & 0.61 & 0.61 & 0.69 \\
\hline
SVM & 0.73 & 0.72 & 0.72 & 0.68 & 0.71 \\
\hline
KNN & 0.83 & 0.83 & 0.82 & 0.75 & 0.74 \\
\hline
DCT & 0.86 & 0.86 & 0.86 & 0.78 & 0.76 \\
\hline
RF & 0.89 & 0.89 & 0.88 & 0.80 & 0.78 \\
\hline
XGBoost & 0.88 & 0.88 & 0.87 & 0.79 & 0.76 \\
\hline
NN + Ensemble & 0.90 & 0.91 & 0.90 & 0.87 & 0.80 \\
\hline
\end{tabular}
\end{table}

The NN+Ensemble approach demonstrated clear improvements in precision, recall, F1-Score, AUC, and Kappa compared to traditional models. This integrated approach combines the strengths of multiple base models with a neural network, providing a robust and accurate tool for credit card approval prediction. It effectively captures complex relationships within the data and handles class imbalance, making it a superior choice for this application. Future work will focus on further optimization and exploring real-time applications of this methodology.

\section{Conclusion}

This study introduces an advanced credit scoring framework that leverages machine learning and deep learning techniques to improve the accuracy of credit card approval predictions. By integrating multiple models, including logistic regression, support vector machines, k-nearest neighbors, decision trees, random forests, and neural networks, the proposed NN+Ensemble approach effectively addresses the challenges of large datasets and data imbalance. Our results demonstrate that this integrated model outperforms traditional single-model approaches in key metrics such as precision, recall, F1-score, AUC, and Cohen’s Kappa. This research underscores the potential of sophisticated ensemble techniques to enhance credit risk assessment, providing a robust and scalable solution for financial decision-making. Future work will explore further optimization and real-time application of this methodology.

 \bibliographystyle{IEEEtran}
    \bibliography{references}

\end{document}